\documentclass[aps,pra,showkeys,twocolumn,superscriptaddress]{revtex4-2}
\usepackage{amsmath, amssymb}
\usepackage{mathrsfs}
\usepackage{array}
\usepackage{booktabs}
\usepackage{tabu}
\usepackage{dcolumn}
\usepackage{amsmath}
\usepackage{amsfonts}
\usepackage{float}
\usepackage{amssymb}
\usepackage{graphicx,color}
\usepackage[colorlinks={true}]{hyperref}
\hypersetup{colorlinks=true,linkcolor=red,citecolor=blue,urlcolor=blue}
\usepackage{graphicx}
\usepackage{subfigure}
\usepackage{graphicx}
\usepackage{dcolumn}
\usepackage{bm}
\usepackage{pstricks}
\usepackage{braket}
\usepackage{orcidlink}

\def\be{\begin{equation}}
  \def\ee{\end{equation}}
\def\bea{\begin{eqnarray}}
\def\eea{\end{eqnarray}}
\def\f{\frac}
\def\n{\nonumber}
\def\l{\label}
\def\p{\phi}
\def\o{\over}
\def\R{\rho}
\def\pa{\partial}
\def\om{\omega}
\def\na{\nabla}
\def\P{\Phi}
\begin{document}
\title{
{ A Study on Thermal Quantum Resources and Probabilistic Teleportation in Spin-1/2 Heisenberg XYZ+DM+KSEA Model under Variable Zeeman Splitting}}

\author{Asad Ali\orcidlink{0000-0001-9243-417X}} \email{asal68826@hbku.edu.qa}
\affiliation{Qatar Centre for Quantum Computing, College of Science and Engineering, Hamad Bin Khalifa University, Doha, Qatar}

\author{Saif Al-Kuwari\orcidlink{0000-0002-4402-7710}}
\affiliation{Qatar Centre for Quantum Computing, College of Science and Engineering, Hamad Bin Khalifa University, Doha, Qatar}

\author{M. T. Rahim\orcidlink{0000-0003-1529-928X}}
\affiliation{Qatar Centre for Quantum Computing, College of Science and Engineering, Hamad Bin Khalifa University, Doha, Qatar}

\author{Mehrdad Ghominejad\orcidlink{0000-0002-0136-7838}}
\affiliation{Faculty of Physics, Semnan University, P.O. Box 35195-363, Semnan, Iran}

\author{Hazrat Ali\orcidlink{0000-0003-1957-3629}}
\affiliation{Department of Physics, Abbottabad University of Science and Technology, P.O. Box 22500 Havellian KP, Pakistan}

\author{Saeed Haddadi\orcidlink{0000-0002-1596-0763}} \email{haddadi@semnan.ac.ir}
\affiliation{Faculty of Physics, Semnan University, P.O. Box 35195-363, Semnan, Iran}
\date{\today}
\def\be{\begin{equation}}
  \def\ee{\end{equation}}
\def\bea{\begin{eqnarray}}
\def\eea{\end{eqnarray}}
\def\f{\frac}
\def\n{\nonumber}
\def\l{\label}
\def\p{\phi}
\def\o{\over}
\def\R{\rho}
\def\pa{\partial}
\def\om{\omega}
\def\na{\nabla}
\def\P{$\Phi$}

\begin{abstract}
We investigate the behavior of various measures of quantum coherence and quantum correlation in the spin-1/2 Heisenberg XYZ model with added Dzyaloshinsky-Moriya (DM) and Kaplan--Shekhtman--Entin-Wohlman--Aharony (KSEA) interactions at a thermal regime described by a Gibbs density operator.
{ We aim to understand the restricted hierarchical classification of different quantum resources, where Bell nonlocality $\subseteq$ quantum steering $\subseteq$ quantum entanglement $\subseteq$ quantum discord $\subseteq$ quantum coherence. This hierarchy highlights the increasingly stringent conditions required as we move from quantum coherence to more specific quantum phenomena.
} In order to enhance quantum coherence, quantum correlation, and fidelity of teleportation, our analysis encompasses the effects of independently provided sinusoidal magnetic field control as well as DM and KSEA interactions on the considered system.
The results reveal that enhancing the entanglement or quantum correlation of the channel does not always guarantee successful teleportation or even an improvement in teleportation fidelity. Thus, the relationship between teleportation fidelity and the channel's underlying quantum properties is intricate.
Our study provides valuable insights into the complex interplay of quantum coherence and correlation hierarchy, offering potential applications for quantum communication and information processing technologies.
\end{abstract}


\maketitle
\section{INTRODUCTION}\label{sec:2}
{ Quantum coherence (QC) is a broad concept that encompasses all kinds of superpositions within both single and multipartite quantum systems \cite{streltsov2017colloquium,baumgratz2014quantifying}. When the off-diagonal elements of a density matrix are nonzero, it indicates the presence of QC. Baumgratz et al. \cite{baumgratz2014quantifying} introduced a quantum resource theory of quantum coherence, where the $l_1$-norm of coherence effectively quantifies the degree of QC.
In the context of multipartite systems, QC can arise from various forms of separable and nonseparable correlations, such as entanglement and quantum discord (QD) \cite{horodecki2009quantum,wootters1998entanglement, ollivier2001quantum,luo2008quantum,ali2010quantum}. Even within these separable and nonseparable correlations, there are additional types of correlations based on their operational interpretation, including Bell nonlocality (BNL) and quantum steering (QS). Together, these different notions form a hierarchy of correlations that are essential for understanding the nature of different types of quantum resources in quantum information and processing.}

\begin{figure}[!t]
\centering
\includegraphics[width=0.55\textwidth]{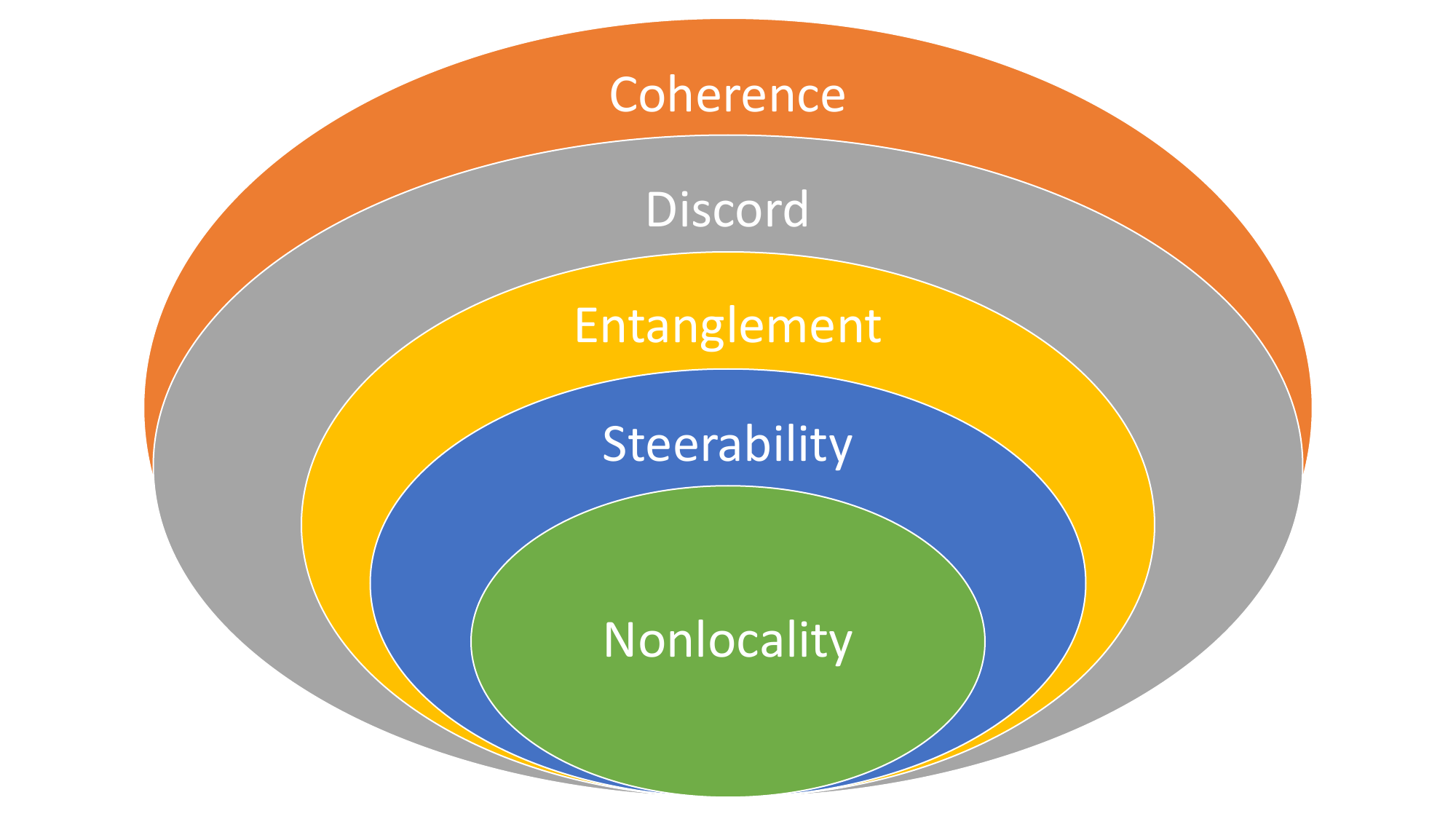}
\caption{ Conventional hierarchy breakdown of quantum coherence and correlations, i.e.,  BNL $\subseteq$ QS $\subseteq$ quantum entanglement $\subseteq$ QD $\subseteq$ QC.}
\label{f1}
\end{figure}

Figure \ref{f1} illustrates the hierarchy of quantum correlations, where each layer represents a subset of the one encompassing it, with quantum coherence being the largest and most inclusive, and BNL being the most specific and restrictive. This hierarchical structure visually represents how each type of quantum correlations may build upon the previous ones, showcasing their interrelations and dependencies \cite{Ma_2019,wiseman2007steering,HaddadiANDP2023}.

Entanglement refers to the remarkable nonseparable quantum correlations between particles, where the state of one particle instantaneously influences the state of another, regardless of distance \cite{Czerwinski1,Czerwinski2,Czerwinski3}. QD  extends beyond entanglement, encompassing { some} quantum correlations residing in both separable and nonseparable states shared between subsystems of a quantum system \cite{ollivier2001quantum,luo2008quantum,ali2010quantum}. { However, in general, it has been shown that quantum correlations may be detected even in the absence of both entanglement and QD \cite{PhysRevLett.126.170404}.} For pure nonseparable states, QD and entanglement are equivalent \cite{ali2010quantum,luo2008quantum}, but this is not the case for mixed states.

QS \cite{schrodinger1935discussion,uola2020quantum,du2021relationship,Haddadi2024qinp} represents a unique form of quantum correlations. In a scenario where two distant observers share an entangled quantum state, QS describes the ability of one observer to remotely influence the state of the other observer's system. While entanglement entails quantum correlations between quantum particles regardless of distance, QS introduces an asymmetry in the observer's roles. In a QS scenario, one observer can manipulate the state of the other observer's system, but the reverse may not always hold. This fundamental asymmetry sets QS apart and raises intriguing questions about the relationship between QS, entanglement, and BNL.

BNL is a concept that arises from the violation of certain inequalities known as Bell inequalities \cite{brunner2014bell,quintino2015inequivalence}. These inequalities describe limits on quantum correlations that can occur between distant systems within a classical framework. When these inequalities are violated, it implies that the quantum correlations between the systems cannot be explained by local hidden variables (underlying properties or information that determine the outcome of measurements)and suggests a form of BNL in quantum mechanics. From Bell inequalities, we now understand that entanglement and BNL are synonymous in a pure state, meaning that entanglement implies BNL and vice versa. However, the situation is not clear and simple for mixed states, even in two-qubit systems \cite{bennett1999quantum,halder2019strong,bhattacharya2020nonlocality,niset2006multipartite}.

{ The measures of BNL, entanglement, QD, and QS are intricately connected, making it difficult to clearly differentiate them, even in systems with two or three qubits. The complexity is underscored by the requirement of 15 independent parameters to fully characterize a density matrix, even for two-qubit mixed state systems, highlighting the intricate nature of quantum correlations.
While entanglement is a cornerstone of quantum information theory, it does not fully encapsulate the correlations present in mixed states. The lack of a straightforward mathematical relationship among these measures, combined with the challenges of distinguishing separable states from entangled ones, indicates that enhancing one type of correlation does not necessarily lead to improvements in others. This complexity necessitates the assessment of multiple metrics, as different quantum resources may become more significant under varying conditions.
This study qualitatively explores the trends and hierarchy of these correlations within a two-qubit Heisenberg model, providing insights into their interrelations and the circumstances under which each type of correlation may dominate.}

A one-dimensional Heisenberg spin chain is a theoretical model generally used in condensed matter physics to describe a linear array of interacting quantum spins \cite{mahmoudi2017non,yin2022markovian,werlang2010thermal,pinheiro2013x,PhysRevE.108.034106,Khedif2022,Hashem2022,Ali_2024}. Each spin interacts with its nearest neighbors, exhibiting phenomena such as quantum phase transitions and quantum correlations. In quantum technologies, Heisenberg spin chains offer potential for quantum simulation \cite{van2021quantum,cappellaro2007simulations}, computing \cite{meier2003quantum,yu2023simulating}, metrology and sensing \cite{liu2021experimental,rams2018limits,ozaydin2015quantum}, quantum criticality and multipartite entanglement \cite{Sun2024}, and communication \cite{chepuri2023complex,abaach2023long}. Entanglement properties of spin chains can be harnessed for quantum communication protocols, where one can use spin chains as quantum channels to transfer quantum states between distant parties securely and with higher fidelity compared to classical communication channels \cite{Benabdallah2022}.

In this work, we take the most general two-qubit Heisenberg XYZ model under a sinusoidally controlled magnetic field applied to an individual spin site with asymmetric spin--orbit coupling interaction called Dzyaloshinsky-Moriya (DM) interaction and symmetric exchange coupling known as Kaplan--Shekhtman--Entin-Wohlman--Aharony (KSEA) interaction, and study the behavior of all the aforementioned quantifiers with changes in the model parameters.

To elucidate the occurrence of weak ferromagnetism observed in certain rhombohedral antiferromagnets, Dzyaloshinsky \cite{dzyaloshinsky1958thermodynamic,dzialoshinskii1957thermodynamic,fert2023early} developed a phenomenological approach grounded in the Landau theory of second-order phase transitions. This approach highlighted that the appearance of a nonzero net magnetization in the system is attributed to the antisymmetric mixed term in the expansion of the thermodynamic potential. Additionally, Dzyaloshinsky noticed that in antiferromagnetic crystals with tetragonal lattices, weak ferromagnetism can be induced by the symmetric mixed term in the corresponding thermodynamic potential. In 1960, Moriya \cite{moriya1960anisotropic} improved the understanding by formulating a microscopic theory of anisotropic superexchange interaction, extending the Anderson theory of superexchange to include spin--orbit coupling. Through perturbation theory, Moriya identified the primary anisotropy contribution to the interaction between neighboring spins as the DM interaction. Moriya also revealed a second-order correction term involving a symmetric traceless tensor. While historically considered negligible, Kaplan and subsequently Shekhtman \emph{et al.}  \cite{shekhtman1993bond,yildirim1995anisotropic,moriya1960anisotropic} argued for the significance of the symmetric term. This term, called KSEA interaction as mentioned before, can restore the O(3) invariance of the isotropic Heisenberg system, a property disrupted by the DM term. This reevaluation of the interaction emphasizes its importance in understanding weak ferromagnetism in antiferromagnetic systems.

Finally, we also consider a teleportation protocol by employing the two-qubit Heisenberg spin chain as a channel to study how the channel parameters can impact the fidelity of the teleported state.
Heisenberg XYZ model has already been experimentally studied, such as in Rydberg or dipolar atoms, through a
combination of dipole interactions with engineered optical
pumping \cite{lee2013unconventional}. Floquet-engineered XXZ spin coupling
in bulk systems and optical tweezer arrays have been
demonstrated in \cite{signoles2021glassy, scholl2022microwave}. Similarly, this model has been recently realized in multiple experiments, notably in dipolar-octupolar (DO) Kramers compounds Ce$_2$(Sn, Zr)$_2$O$_7$ and Nd$_2$Zr$_2$O$_7$ \cite{gaudet2019quantum, patri2020distinguishing, chen2023coulombic}. Moreover, the XYZ model has been realized in Weyl-Heisenberg ferromagnets \cite{rosenberg2023quantum}.

{ \subsection{Motivation and contribution of this study}

This study aims to enhance our understanding of various quantum resources in quantum information processing, particularly in quantum teleportation. While entanglement is vital for teleportation, the complexities of mixed states affected by thermal and quantum noise necessitate the consideration of additional quantum correlations, such as BNL, QS, QD, and $l_1$-norm of QC. Utilizing the Heisenberg XYZ spin model, we explore how the Gibbs thermal state acts as a mixed state channel for teleportation, incorporating various control parameters like magnetic fields and interactions such as DM and KSEA. This approach allows us to investigate how these factors influence the different coherence and correlation measures constituting the hierarchy of quantum resources and the fidelity of teleportation.
Our contribution lies in systematically quantifying these quantum resources and analyzing their collective impact on teleportation fidelity. We also examine the robustness of these resources under varying temperature conditions, providing insights into optimizing quantum communication protocols.
}

\subsection{Organization}
The scheme of this paper is as follows: in Sec. \ref{sec2}, we define and explain the different notions of quantum coherence and quantum correlation measures.  In Sec. \ref{sec3}, we briefly introduce the spin-1/2 Heisenberg XYZ model, diagonalize the system, and come up with the thermal state of the model. In Sec. \ref{sec4}, we present a simple teleportation scheme based on the considered model to be implemented.
Section \ref{sec5} presents the results and discussion, and finally, Sec. \ref{sec6} concludes this paper.

\section{Quantum resources}\label{sec2}

\subsection{$l_1$-norm quantum coherence}
Quantum coherence is a fundamental concept that arises from the superposition principle in quantum mechanics. A rigorous framework to quantify coherence as a resource, known as the resource theory of quantum coherence, has been previously developed  \cite{streltsov2017colloquium,baumgratz2014quantifying}. This theory identifies the set of incoherent states $\mathcal{I}$ which are diagonal in a reference basis $\{|i\rangle\}$:
\begin{equation}
\delta \in \mathcal{I} \iff \delta = \sum_i \delta_i |i\rangle\langle i|.
\label{eq1}
\end{equation}
The free operations are the incoherent operations that map incoherent states to incoherent states. Revealing and quantifying quantum coherence is essential to enable quantum correlations and information processing.

Accordingly, Baumgratz \emph{et al.} \cite{baumgratz2014quantifying} proposed the $l_1$-norm of quantum coherence as a quantifier of coherence, given by
\begin{equation}
Q(\rho) = \sum_{i\neq j} |\langle i|\rho|j\rangle| = \sum_{i,j} |\rho_{ij}| - \sum_i |\rho_{ii}|,  \label{eq2}
\end{equation}
where $\rho$ is the density operator of the considered system.

\subsection{Quantum discord}
QD measures nonclassical correlations between subsystems in a quantum system, capturing quantumness not explained by classical means. QD quantifies differences between quantum and classical mutual information.

QD indicates the difference between total and classical correlation for a two-qubit system \cite{ali2010quantum,ollivier2001quantum,luo2008quantum,fanchini2010non}. That is, QD for a bipartite quantum X state, $\rho_{X}$ is expressed as
\begin{equation}
D(\rho_{X}) = \min\{q_1,~q_2\},
\label{eq3}
\end{equation}
where
\begin{equation}
q_j = H(\rho_{11}+\rho_{33}) + \sum^4_{i=1} \lambda_i \log_2 \lambda_i + w_j,
\label{eq4}
\end{equation}
here, $\lambda_i$'s represent the eigenvalues of density matrix $\rho_{X}$, and
\begin{equation}
w_1=H(\xi),
\label{eq5}
\end{equation}
\begin{equation}
w_2=-\sum_i \rho_{ii} \log_2 \rho_{ii} - H(\rho_{11}+\rho_{33}),
\label{eq6}
\end{equation}
with
\begin{equation}
H(\xi)=-\xi \log_2 \xi - (1-\xi)\log_2(1-\xi),
\label{eq7}
\end{equation}
and
\begin{equation}
\xi=\frac{1}{2}\left\{1+\sqrt{[1-2(\rho_{33}+\rho_{44})]^2+4(|\rho_{14}|+|\rho_{23}|)^2}\right\}.
\label{eq8}
\end{equation}

\subsection{Concurrence}
Concurrence is an entanglement monotone used to quantify the degree of entanglement in arbitrary two-qubit states \cite{wootters1998entanglement,wootters2001entanglement}. For a given density matrix $\rho$, concurrence $C(\rho)$ of two qubits is defined as follows
\begin{equation}
C(\rho) = \max\{0, \lambda_1 - \lambda_2 - \lambda_3 - \lambda_4\}.
\label{eq13}
\end{equation}
where $\lambda_i$'s are the square roots of the eigenvalues of the matrix (non-Hermitian) $\rho (\sigma_y \otimes \sigma_y) \rho^* (\sigma_y \otimes \sigma_y)$ in decreasing order. As before, $\sigma_y$ represents the $y$-component of Pauli matrices, and $\rho^*$ denotes the complex conjugate of $\rho$. Note that the measure \eqref{eq13} takes values between 0 and 1, indicating the absence of entanglement when $C(\rho) = 0$ and maximal entanglement when $C(\rho) = 1$.

\subsection{Quantum steering}
Erwin Schrödinger proposed steering as an extension of the Einstein-Podolsky-Rosen paradox \cite{schrodinger1935discussion}. Though initially overlooked as nonlocal correlations, it later sparked significant advancements. Steerable states exhibit quantum advantages in device-independent quantum cryptography \cite{branciard2012one}, secure teleportation \cite{he2015secure}, randomness generation \cite{law2014quantum}, and subchannel discrimination \cite{piani2015necessary}. The three-setting linear steering inequality \cite{cavalcanti2009experimental,costa2016quantification} is based on the assumption that either Alice or Bob are allowed three measurement observables on their respective subsystems. It  is a crucial tool for detecting and quantifying steering in a state, particularly in a two-qubit system which reads
\begin{equation}
\frac{1}{\sqrt{3}} \sum_{i=1}^3 \mathrm{Tr}(A_i \otimes B_i\rho) \leq 1,
\label{eq14}
\end{equation}
where $A_i = a_i \cdot \sigma$ and $B_i = b_i \cdot \sigma$ are Hermitian operators acting on qubits $A$ and $B$, respectively. Here, $a_i, b_i \in \mathbb{R}^3$ are two unit vectors with $\{b_1, b_2, b_3\}$ being orthogonal to each other, and Pauli matrices $\sigma = (\sigma_1=\sigma_x, \sigma_2=\sigma_y, \sigma_3=\sigma_z)$. Any violation of the inequality \eqref{eq14} implies that $\rho$ is steerable, and the maximal violation gives a measure of steering  \cite{costa2016quantification}.  Explicitly,
\begin{equation}
S(\rho) = \max_{\{A_i,B_i\}} \left \{ \frac{1}{\sqrt{3}} \sum_{i=1}^3 \mathrm{Tr}(A_i \otimes B_i\rho) \right \}
\label{eq15}
\end{equation}
represents the maximum violation of the three-setting linear steering inequality.

\subsection{Bell nonlocality}
In order to quantify BNL and enable a more thorough comparison of quantum correlations, one can use the Bell inequality violation. The normalized form of BNL measure can be written as \cite{bartkiewicz2013entanglement,bartkiewicz2017bell,horodecki1995violating,sun2017relativistic,hu2013relations}
\begin{equation}
\mathcal{B}(\rho)= \max\left\{0, \frac{\mathcal{B}_{\textmd{CHSH}}-2}{\mathcal{B}_{\text{max}}-2}\right\},
\label{eq16}
\end{equation}
where $0\leq\mathcal{B}(\rho)\leq1$ since $\mathcal{B}_{\textmd{CHSH}}\leq \mathcal{B}_{\text{max}}=2\sqrt{2}$ for a two-qubit system $\rho$. { Notice,
$\mathcal{B}_{\textmd{CHSH}}$ is the maximum violation of Bell Clauser--Horne--Shimony--Holt (CHSH) inequality, which for an X-shaped density matrix is given by \cite{rahman2023two,sun2017relativistic,hu2013relations}:
{\small
\begin{align}
&\mathcal{B}_{\textmd{CHSH}} = 2\max \Big\{ 2\sqrt{2|\rho_{14}|^2 + 2|\rho_{23}|^2},
 \sqrt{4(|\rho_{14}| + |\rho_{23}|)^2 +  \gamma^2} \Big\},
\end{align}}
with
$\gamma=\rho_{11} - \rho_{22} - \rho_{33} + \rho_{44}$.}
\section{Heisenberg XYZ model}\label{sec3}
The Hamiltonian describing a general Heisenberg XYZ model under the influence of an inhomogeneous magnetic field, including DM and KSEA interactions, is given by
\begin{equation}
\hat{\mathcal{H}}_N=\hat{\mathcal{H}}_H+\hat{\mathcal{H}}_B+\hat{\mathcal{H}}_{\textmd{DM}}+\hat{\mathcal{H}}_{\textmd{KSEA}},
\label{eq17}
\end{equation}
The initial term $\hat{\mathcal{H}}_H$ accounts for Heisenberg spin exchange interactions, with $\hat{\mathcal{H}}_B$ as the Zeeman Hamiltonian representing the influence of the inhomogeneous external magnetic field on the system. Additionally, $\hat{\mathcal{H}}_{\textmd{DM}}$ and $\hat{\mathcal{H}}_{\textmd{KSEA}}$ represent the DM interaction and KSEA interaction, respectively.
This total Hamiltonian \eqref{eq17} can be explicitly written as \cite{Yurischev2020,Yurischev2023}
\begin{equation}
\begin{split}
\hat{\mathcal{H}}_N =& \sum_{i=1}^{N-1}\big\{
 J_x \hat{\sigma}_i^x\otimes\hat{\sigma}_{i+1}^x + J_y \hat{\sigma}_{i}^y\otimes\hat{\sigma}_{i+1}^y + J_z \hat{\sigma}_{i}^z\otimes\hat{\sigma}_{i+1}^z \\
& +\overrightarrow{\mathbf{B}}_{i}\cdot(\overrightarrow{\mathbf{\mathcal{\sigma}}}_{i}\otimes\mathbb{I}_2)+\overrightarrow{\mathbf{B}}_{i+1}\cdot(\mathbb{I}_2\otimes\overrightarrow{\mathcal{\sigma}}_{i+1})\\
& + \overrightarrow{\mathbf{\mathcal{D}}}\cdot(\overrightarrow{\mathbf{\mathcal{\sigma}}}_{i}\times\overrightarrow{\mathbf{\mathcal{\sigma}}}_{i+1})+  \overrightarrow{\mathbf{\mathcal{\sigma}}}_{i}\cdot\Gamma\cdot\overrightarrow{\mathbf{\mathcal{\sigma}}}_{i+1}\big\},
\end{split}
\label{eq18}
\end{equation}
here,
\begin{equation}
\Gamma = \begin{pmatrix}
0 & \Gamma_{z} & \Gamma_{y} \\
\Gamma_{z} & 0 & \Gamma_{x} \\
\Gamma_{y} & \Gamma_{x} & 0 \\
\end{pmatrix}.
\label{eq19}
\end{equation}

The interaction between neighboring spin sites is determined by the real spin--spin coupling coefficients along the $x$, $y$, and $z$ directions, denoted as $J_k$ for $k = x, y, z$. Antiferromagnetic coupling is indicated by positive values of $J_k$, while negative values represent ferromagnetic coupling. The standard Pauli spin operators, denoted as $\hat{\sigma}_i^k$, define the spin vector on the $i$th site as $\overrightarrow{\mathbf{\mathcal{\sigma}}}_{i} = (\hat{\sigma}_i^x, \hat{\sigma}_i^y, \hat{\sigma}_i^z)$. Moreover,
\(\overrightarrow{\mathbf{B}}_{i}=(B_{i}^x,B_{i}^y,B_{i}^z)\) represents the magnetic flux density contributing to Zeeman splitting at the $i$th spin site.
The DM interaction vector is defined by \(\overrightarrow{\mathbf{\mathcal{D}}}=(\mathcal{D}_x,\mathcal{D}_y,\mathcal{D}_z)\) and serves as the anti-symmetric exchange coupling responsible for spin--orbit coupling in specific materials. Note that the KSEA interaction ($\Gamma$) introduces an anisotropic symmetric interaction \cite{Yurischev2020,Yurischev2023,Khedif2021,Kuznetsova2023}.

{ To simplify our analysis, we focus on a bipartite Heisenberg XYZ model restricted to interactions between two specific spin sites. This model is limited to a two-spin system and does not consider interactions between arbitrary pairs within a longer chain or under periodic boundary conditions. For this two-spin model, we further simplify by assuming that $\overrightarrow{\mathbf{B}}_{1} = (0,0,B\cos{\theta})$, $\overrightarrow{\mathbf{B}}_{2} = (0,0,B\sin{\theta})$, $\overrightarrow{\mathbf{\mathcal{D}}} = (0,0,\mathcal{D}_z)$, and specifying that $\Gamma_x = \Gamma_y = 0$ while $\Gamma_z \neq 0$. Consequently, the Hamiltonian in Eq. \eqref{eq18} with these assumptions can be explicitly written as
}
\begin{equation}
\begin{split}
\hat{\mathcal{H}} = &
 J_x~ \hat{\sigma}_1^x\otimes\hat{\sigma}_{2}^x + J_y~ \hat{\sigma}_{1}^y\otimes\hat{\sigma}_{2}^y + J_z~ \hat{\sigma}_{1}^z\otimes\hat{\sigma}_{2}^z \\
& + B~[\cos{\theta}(\hat{\sigma}_{1}^z\otimes\mathbb{I}_2) + \sin{\theta}(\mathbb{I}_2\otimes\hat{\sigma}_{2}^{z})] \\
& +  \mathcal{D}_z~ (\hat{\sigma}_1^x\otimes\hat{\sigma}_{2}^y - \hat{\sigma}_{1}^y\otimes\hat{\sigma}_{2}^x)\\
& +  \Gamma_z~ (\hat{\sigma}_1^x\otimes\hat{\sigma}_{2}^y + \hat{\sigma}_{1}^y\otimes\hat{\sigma}_{2}^x).
\end{split}
\label{eq20}
\end{equation}

As we know, this Hamiltonian \eqref{eq20} has a standard X-structured matrix form. Using the standard computational basis $\{\ket{00}, \ket{01}, \ket{10}, \ket{11}\}$, it can be expressed in a simplified form
\begin{widetext}
\begin{align}
\hat{\mathcal{H}} =
\begin{pmatrix}
J_z+B(\cos{\theta}+\sin{\theta}) & 0 & 0 & -2 i \Gamma_z +J_x-J_y \\
 0 & -J_z+B(\cos{\theta}-\sin{\theta}) & 2 i \mathcal{D}_z+J_x+J_y & 0 \\
 0 & -2 i \mathcal{D}_z+J_x+J_y & -J_z-B(\cos{\theta}-\sin{\theta}) & 0 \\
 2 i \Gamma_z +J_x-J_y & 0 & 0 & J_z-B(\cos{\theta}+\sin{\theta}) \label{eq21}
\end{pmatrix}.
\end{align}
\end{widetext}

The diagonalization of $\hat{\mathcal{H}}$ \eqref{eq21} results in four eigenvalues, given by
\begin{equation}
\begin{aligned}
E_{1,2} &= \pm{\mathcal{K}_1} + J_z, \\
E_{3,4} &= \pm{\mathcal{K}_2} - J_z,
\end{aligned}
\label{eq22}
\end{equation}
with expressions
$$
\mathcal{K}_1=\sqrt{k_1^2+B^2\left(\cos{\theta}+\sin{\theta}\right)^2}$$
and
$$\mathcal{K}_2=\sqrt{k_2^2+B^2\left(\cos{\theta}-\sin{\theta}\right)^2},$$
where
$$
k_1=\sqrt{\left(J_x-J_y\right)^2+4 \Gamma_z^2}$$
and
$$k_2=\sqrt{\left(J_x+J_y\right)^2+4 \mathcal{D}_z^2}.$$

The state of a system in the thermal equilibrium with the absolute temperature $T$ is described by the Gibbs density operator
\begin{equation}
\rho_T = \frac{1}{\mathcal{Z}} e^{-\hat{\mathcal{H}}/\kappa T},
\label{eq31}
\end{equation}
where $\mathcal{Z}=\mathrm{Tr}[\exp(-\hat{\mathcal{H}}/\kappa T]$ is partition function with the Boltzmann constant $\kappa$ (considered as $\kappa=1$ for simplification purpose).

Using now Hamiltonian \eqref{eq21} and density operator \eqref{eq31}, we obtain the following density matrix at thermal equilibrium

\begin{equation}
\rho_T=\left(\begin{array}{cccc}
a^- & 0 & 0 & c \\
0 & b^- & d & 0 \\
0 & d^* & b^+ & 0 \\
c^* & 0 & 0 & a^+
\end{array}\right),\label{eq33}
\end{equation}
where the nonzero elements read
\begin{equation}
\begin{aligned}
&a^{\pm}  =\frac{e^{- J_z/T}}{\mathcal{Z}}\left\{\cosh \left( \mathcal{K}_1/T\right)\pm\left[B(\cos{\theta}+\sin{\theta}) / \mathcal{K}_1\right] \Theta_1\right\} , \\
&b^{\pm}  =\frac{e^{J_z/T}}{\mathcal{Z}}\left\{\cosh \left( \mathcal{K}_2/T\right)\pm\left[B(\cos{\theta}-\sin{\theta}) / \mathcal{K}_2\right] \Theta_2\right\}, \\
&c =-\frac{e^{- J_z/T}}{\mathcal{Z}}\left[\left(-2 i \Gamma_z+J_x-J_y\right) / \mathcal{K}_1\right] \Theta_1 , \\
&d  =-\frac{e^{J_z/T}}{\mathcal{Z}}\left[\left(2 i \mathcal{D}_z+J_x+J_y\right) / \mathcal{K}_2\right] \Theta_2,\label{elements}
\end{aligned}
\end{equation}
where $\Theta_{(1)2}=\sinh \left( \mathcal{K}_{(1)2}/T\right)$, and
\begin{equation}
Z=2\left[e^{- J_z/T} \cosh \left( \mathcal{K}_1/T\right)+e^{ J_z/T} \cosh \left(\mathcal{K}_2/T\right)\right],\label{eq32}
\end{equation}

The state presented in Eq. \eqref{eq33} has an X structure. Thus, the explicit expressions of the aforementioned quantifiers based on thermal state elements \eqref{elements} can be derived straightforwardly.

For example, the $l_1$-norm of quantum coherence \eqref{eq2} can be obtained as
\begin{equation}
Q(\rho_T)=|c|+|c^*|+|d|+|d^*|.
\label{eq35}
\end{equation}

Moreover, the analytical expression of QD for our thermal state \eqref{eq33} would be given by Eq. \eqref{eq3} with considering $\rho_{11}=a^-$, $\rho_{22}=b^-$, $\rho_{33}=b^+$, $\rho_{44}=a^+$, $\rho_{14}=c$, and $\rho_{23}=d$.

Regarding the analytical form of QS \eqref{eq15}, one needs to simply obtain $\mathrm{Tr}(A_i \otimes B_i\rho_T)$. Explicitly, we have
\begin{equation}
\begin{split}
S(\rho_{T}) = & \big\{8 (c c^*+d d^*) - 4 a^- (b^- + b^+) \\
& - 4 a^+ (b^- - b^+) + 1\big\}^{1/2}.
\end{split}
\label{eq38}
\end{equation}

Besides, the concurrence \eqref{eq13} can be derived as
\begin{equation}
C(\rho_{T})=2\max \left\{0, t_1, t_2\right\},
\end{equation}
where $ t_1=|d|-\sqrt{a^- a^+}$ and $t_2=|c|-\sqrt{b^- b^+}$.

Finally, the normalized
form of BNL function \eqref{eq16} is found to be
\begin{equation}
\mathcal{B}(\rho_T)= \max\left\{0, (2  \max\{m_1, m_2\}-2)/(2\sqrt{2}-2)\right\},
\end{equation}
where $m_1=\sqrt{4(|c| + |d|)^2 + (a^- - b^- - b^+ + a^+)^2}$ and
$m_2=\sqrt{8(|c|^2 + |d|^2)}$.

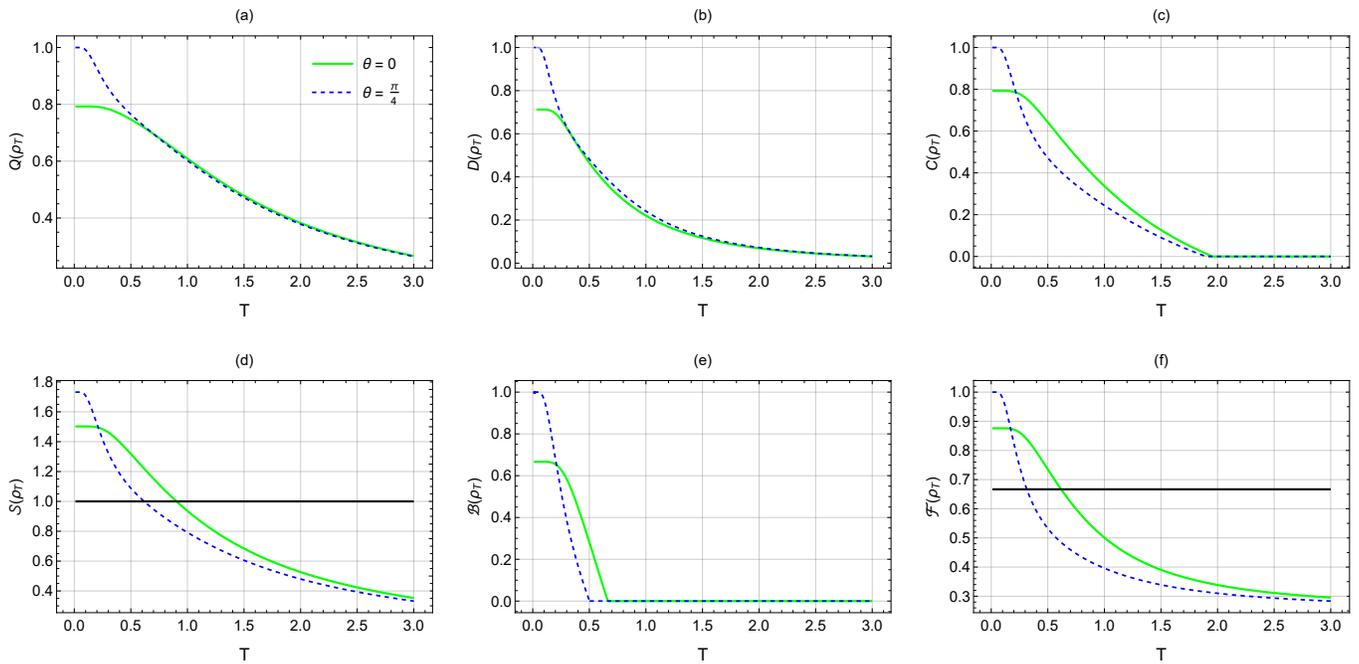
\begin{figure*}[!t]
\centering
\includegraphics[width=0.9\textwidth]{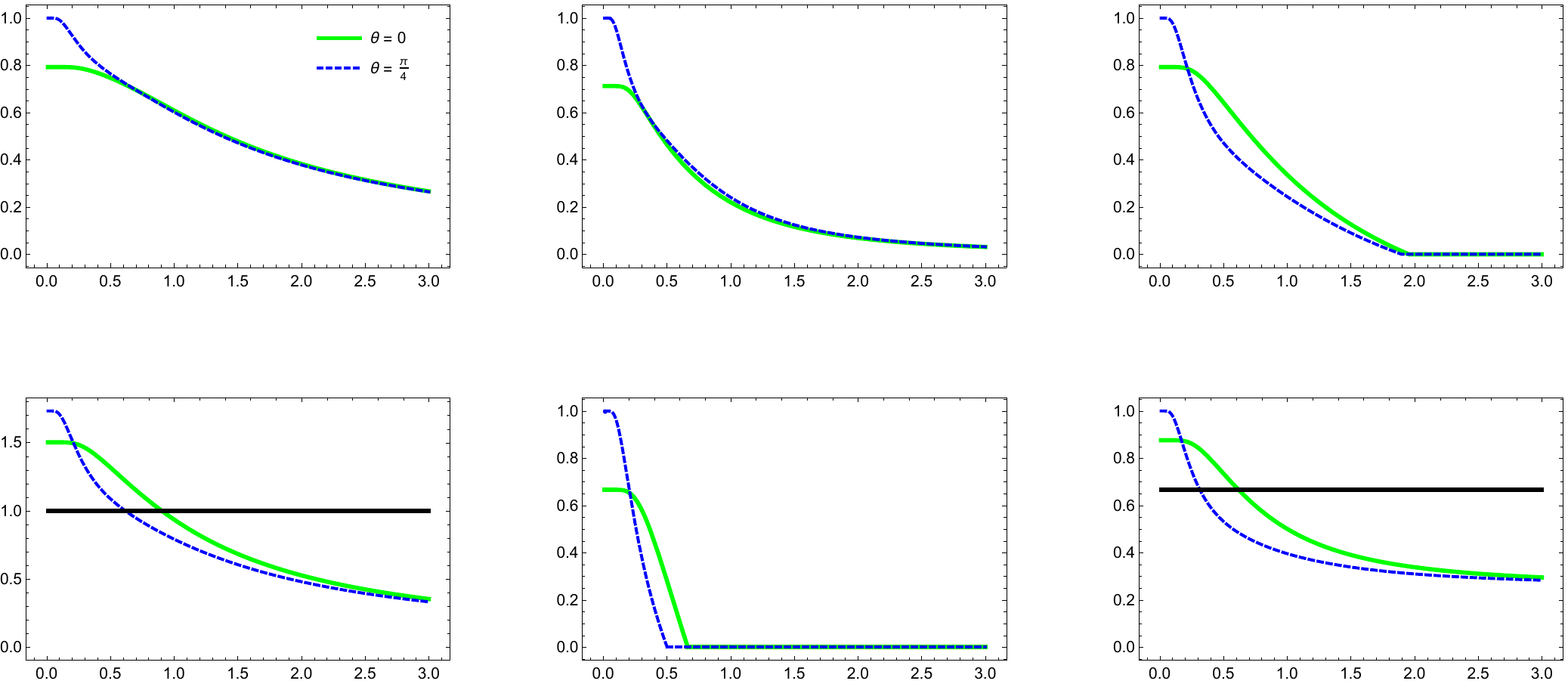}
\put(-445,130){$(\bold{a})$}
\put(-282,130){$(\bold{b})$}
\put(-120,130){$(\bold{c})$}
\put(-445,20){$(\bold{d})$}
\put(-282,20){$(\bold{e})$}
\put(-120,20){$(\bold{f})$}
\put(-488,158){$Q(\rho_T)$}
\put(-323,158){$D(\rho_T)$}
\put(-158,158){$C(\rho_T)$}
\put(-488,47){$\mathcal{S}(\rho_T)$}
\put(-323,47){$\mathcal{B}(\rho_T)$}
\put(-159,47){$\mathcal{F}(\rho_T)$}
\put(-392,-10){$T$}
\put(-392,105){$T$}
\put(-230,-10){$T$}
\put(-230,105){$T$}
\put(-65,-10){$T$}
\put(-65,105){$T$}
\caption{The variations of $l_1$-norm quantum coherence (a), QD (b), concurrence (c), QS (d),  BNL (e), and average fidelity (f) as a function of temperature $T$ at $J_x=0.8$, $J_y=0.5$, $J_z=0.3$, $\mathcal{D}_z=0$, $\Gamma_z=0$ when the magnetic field is applied to single spin ($\theta=0$), and when a same magnetic field is applied to both spins ($\theta=\pi/4$). Black horizontal lines show $S(\rho_T)=1$ in (d) and $\mathcal{F}(\rho_T)=2/3$ in (f).}
\label{f2}
\end{figure*}
\section{Quantum teleportation}\label{sec4}
In this section, we examine the teleportation protocol, considering $\rho_T$ as the quantum channel state, assuming the input state is an arbitrary unknown pure state consisting of two qubits such as
\begin{equation}
|\psi_{\text{in}}\rangle = \cos(\alpha/2)|10\rangle + e^{i\phi} \sin(\alpha/2)|01\rangle,
\label{eq40}
\end{equation}
where $\alpha$ and $\phi$ represent the amplitude and phase of the target state to be teleported, respectively.
The quantum channel is known as a completely positive and trace-preserving operator. Through this process, an input state is mapped to an output state.

When a quantum state is teleported via the mixed channel $\rho_{\text{ch}}$, the resulting output replica state  $\rho_{\text{out}}$ is achieved by performing joint measurements and local unitary transformations on the input state $\rho_{\text{in}}=|\psi_{\text{in}}\rangle\langle\psi_{\text{in}}|$

\begin{equation}
\rho_{\text {out }}=\sum_{i, j \in\{0, x, y, z\}} p_i p_j\left(\sigma^i \otimes \sigma^j\right) \rho_{\text {in }}\left(\sigma^i \otimes \sigma^j\right),
\end{equation}
where $\sigma^0=\mathbb{I}_2$, and $p_i=\operatorname{Tr}\left(q^i \rho_{\text {ch}}\right)$ satisfies the condition $\sum_i p_i=1$. Moreover, we have $q^0=|\Psi^{-}\rangle\langle\Psi^{-}|$, $q^1=| \Phi^{-}\rangle\langle\Phi^{-}|$, $q^2=| \Psi^{+}\rangle\langle\Psi^{+}|$, and $q^3=|\Phi^{+}\rangle\langle\Phi^{+}|$ in which $|\Psi^{ \pm}\rangle$ and $|\Phi^{ \pm}\rangle$ are four Bell states.
Here, let us suppose that the quantum channel is a thermal state \eqref{eq33}, namely $\rho_{\mathrm{ch}}=\rho_T$.

The quality of the teleported state is determined by the criterion of fidelity $F(\rho_{\text{in}}, \rho_{\text{out}})$, defined as
\begin{equation}
F(\rho_{\text{in}}, \rho_{\text{out}}) = \left(\mathrm{Tr}\sqrt{\sqrt{\rho_{\text{in}}}\rho_{\text{out}}\sqrt{\rho_{\text{in}}}}\right)^2.
\end{equation}

Based on the above definition, the average fidelity of teleportation $\mathcal{F}$ is given by
\begin{equation}
\mathcal{F}= \frac{1}{4\pi}\int_{0}^{2\pi} \text{d}\phi \int_{0}^{\pi} \text{d}\alpha F(\rho_{\text{in}}, \rho_{\text{out}})\sin \alpha.
\end{equation}

The maximum classical average fidelity threshold is observed at $\mathcal{F} = 2/3$. Beyond this point, we transition into the quantum average fidelity regime. The proximity of the quantum average fidelity to unity signifies reduced information leakage, indicating optimal conditions for quantum teleportation.

In our case (density matrix \eqref{eq33}), the analytical form of average fidelity is derived by
\begin{equation}
\mathcal{F}(\rho_{T})=\frac{2}{3}(b^- +b^+)^2 +\frac{1}{3}(a^- +a^+)^2+\frac{1}{3}(d +d^*)^2  .
\end{equation}

\section{Results and discussion}\label{sec5}

In this investigation, our objective is to analyze various quantum coherence and correlation measures within the Heisenberg XYZ model, using it as a quantum resource channel for teleporting a target quantum state $|\psi_{\text{in}}\rangle = \cos(\alpha/2)|10\rangle + e^{i\phi} \sin(\alpha/2)|01\rangle$. The goal is to enhance average quantum teleportation fidelity beyond 2/3 by adjusting parameters such as magnetic field, bath temperature, and DM and KSEA interactions, thereby enhancing the channel's quantum resources.
We consider an antiferromagnetic spin chain characterized by anisotropic spin--spin coupling constants (a random case): \( J_x = 0.8 \), \( J_y = 0.5 \), and \( J_z = 0.3 \).
\subsection{Impact of Magnetic Field}\label{subsec1}
First, our focus is on the impact of applying a magnetic field to either one spin (\(\theta = 0\) or \(\theta = \pi/2\)) or to both spins (\(\theta = \pi/4\)) with $B=1$ on quantum resources and teleportation fidelity in the absence of DM and KSEA interactions. Figure \ref{f2} illustrates the variations of quantum resources and teleportation fidelity with temperature under zero DM and KSEA interactions, comparing two distinct fixed magnetic field values.

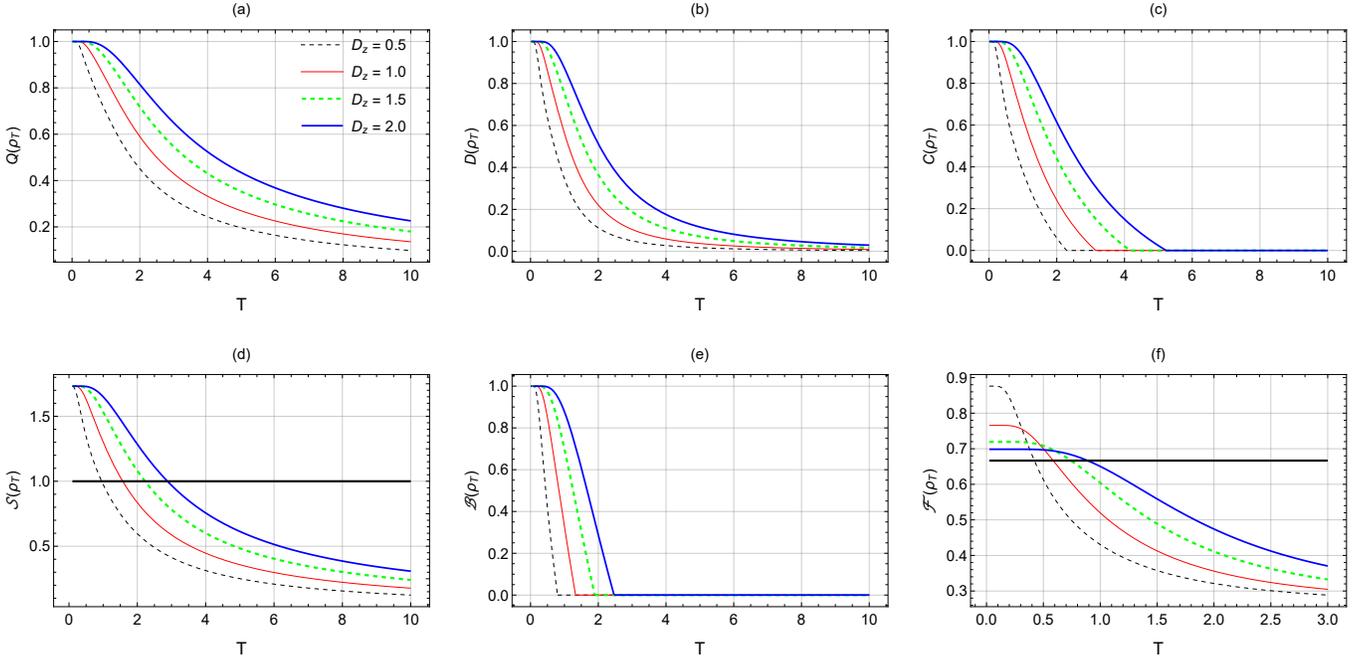
\begin{figure*}[!t]
\centering
\includegraphics[width=0.9\textwidth]{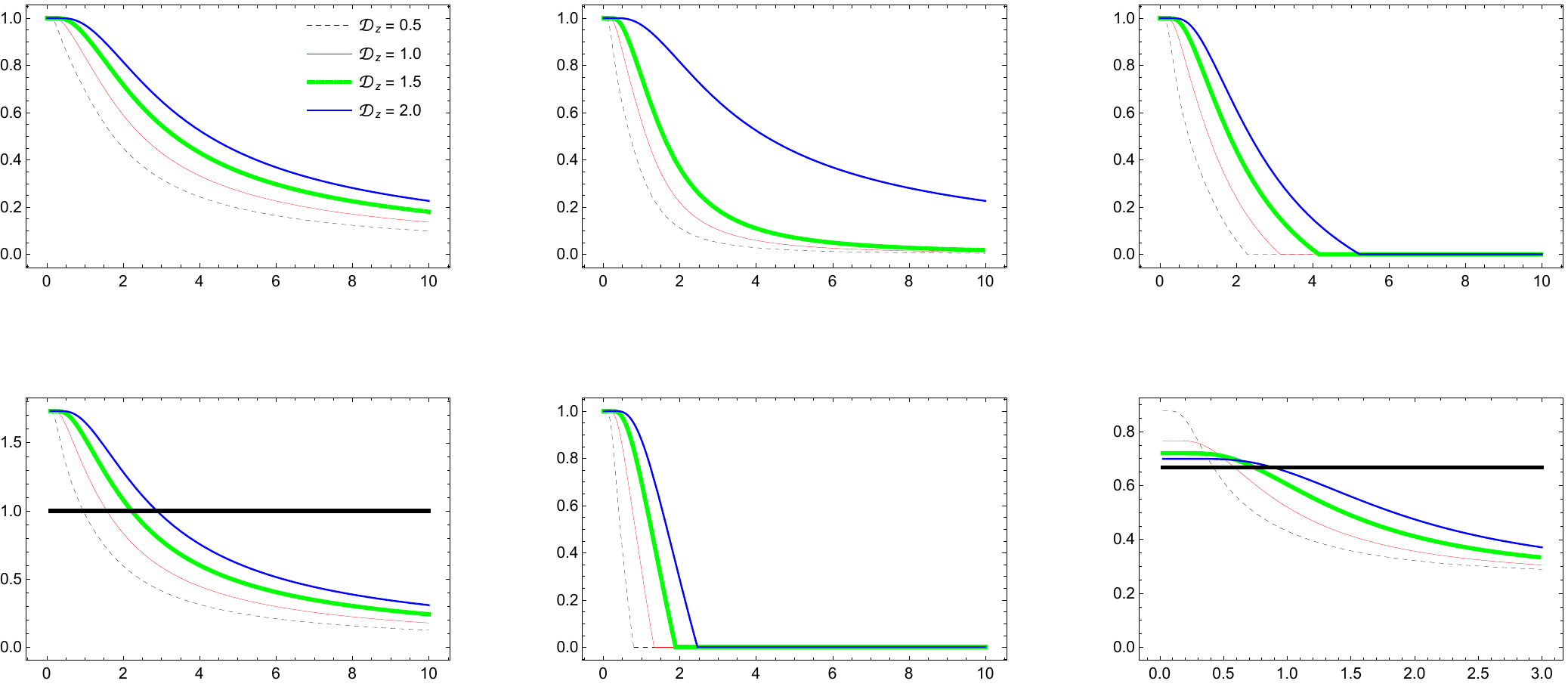}
\put(-445,130){$(\bold{a})$}
\put(-282,130){$(\bold{b})$}
\put(-120,130){$(\bold{c})$}
\put(-442,17){$(\bold{d})$}
\put(-288,17){$(\bold{e})$}
\put(-120,17){$(\bold{f})$}
\put(-488,158){$Q(\rho_T)$}
\put(-323,158){$D(\rho_T)$}
\put(-158,158){$C(\rho_T)$}
\put(-488,47){$\mathcal{S}(\rho_T)$}
\put(-323,47){$\mathcal{B}(\rho_T)$}
\put(-159,47){$\mathcal{F}(\rho_T)$}
\put(-392,-10){$T$}
\put(-392,105){$T$}
\put(-230,-10){$T$}
\put(-230,105){$T$}
\put(-65,-10){$T$}
\put(-65,105){$T$}
\caption{
The behaviors of $l_1$-norm quantum coherence (a), QD (b), concurrence (c), QS (d),  BNL (e), and average fidelity (f) versus temperature $T$ at $J_x=0.8$, $J_y=0.5$, $J_z=0.3$, $\theta=\pi/4$, and $\Gamma_z=0$ for different fixed values of $\mathcal{D}_z$.}
\label{f3}
\end{figure*}

In Figs. \ref{f2}(a) and \ref{f2}(b), the \( l_1 \) norm of quantum coherence and QD are shown as functions of temperature. At low temperatures, peak values are higher when the same magnetic field is applied to both spins (dashed blue curve) compared to when it is applied to just one spin (solid green curve). Both measures decrease with increasing temperature. This result does not surprise us because the temperature can diminish the quantum coherence and correlation due to thermal fluctuations in the system.

Figures \ref{f2}(c), \ref{f2}(d), and \ref{f2}(e) depict the behaviors of concurrence, QS, and  BNL versus $T$, respectively. The entanglement captured by concurrence disappears after $T\approx 1.9$. Note that at low temperatures, peak values of concurrence are greater when the same magnetic field is applied to both spins (dashed blue curve). Our findings show that for \( 0.7 \leq T \leq 0.9 \), the channel state is steerable without BNL, as \( S(\rho_T) > 1 \) in Fig. \ref{f2}(d). Between \( T \approx 0.7 \) and \( T \approx 1.9 \), the channel state remains entangled [see Fig. \ref{f2}(c)] but without BNL [see Fig. \ref{f2}(e)]. More precisely,  BNL decreases to zero before concurrence and discord as the channel becomes thermally mixed. Notice, \( Q(\rho_T) \) and \( D(\rho_T) \) show more robustness compared to \( C(\rho_T) \), \( \mathcal{S}(\rho_T) \), and \( \mathcal{B}(\rho_T) \).  Thus, the mixed entangled states can exist without violating Bell inequalities or exhibiting BNL.
These highlight the hierarchical relationship between the mentioned quantum resources.

Our results in Fig. \ref{f2} indicate that applying the same magnetic field to both qubits significantly enhances average teleportation fidelity [see Fig. \ref{f2}(f)], quantum coherence, and correlations at low temperatures. These enhancements diminish at higher temperatures due to thermal effects { \cite{el2018quantifying,qars2024gaussian}}. However, fidelity remains above the classical limit $\mathcal{F}(\rho_T)=2/3$ at low temperatures for both \(\theta = 0\) and \(\theta = \pi/4\), with maximum fidelity achieved for \(\theta = \pi/4\).

From Fig. \ref{f2}, we conclude that applying the same magnetic field to both qubits enhances teleportation fidelity and quantum correlations more significantly across different temperatures compared to applying it to just one qubit. The hierarchy of quantum resources, i.e.,  BNL $\subseteq$ QS $\subseteq$ quantum entanglement $\subseteq$ QD $\subseteq$ QC depicted in Figs. \ref{f1} and \ref{f2}, demonstrates that the BNL function is the weakest measure, while the \( l_1 \)-norm of coherence is the strongest indicator of nonclassical characteristics.

Given that using the same magnetic field control for both spins enhances quantum coherence and correlations as well as average teleportation fidelity at low temperatures, we will adopt $\theta=\pi/4$ as the standard setting and explore other model parameters to improve these measures against the increasing of temperature.

\begin{figure*}[!t]
\centering
\includegraphics[width=0.9\textwidth]{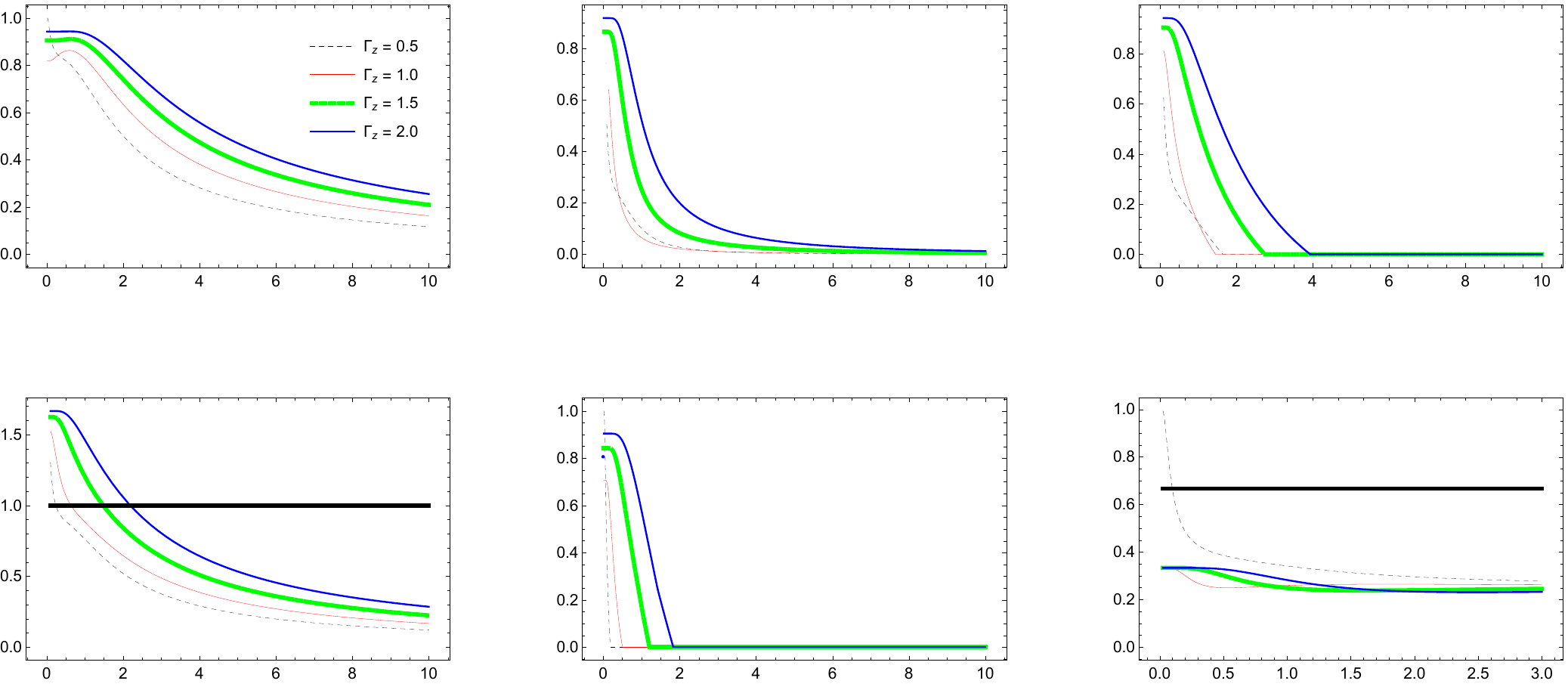}
\put(-445,130){$(\bold{a})$}
\put(-286,128){$(\bold{b})$}
\put(-120,130){$(\bold{c})$}
\put(-445,16){$(\bold{d})$}
\put(-262,16){$(\bold{e})$}
\put(-120,16){$(\bold{f})$}
\put(-488,158){$Q(\rho_T)$}
\put(-323,158){$D(\rho_T)$}
\put(-158,158){$C(\rho_T)$}
\put(-488,47){$\mathcal{S}(\rho_T)$}
\put(-323,47){$\mathcal{B}(\rho_T)$}
\put(-159,47){$\mathcal{F}(\rho_T)$}
\put(-392,-10){$T$}
\put(-392,105){$T$}
\put(-230,-10){$T$}
\put(-230,105){$T$}
\put(-65,-10){$T$}
\put(-65,105){$T$}
\caption{The behaviors of $l_1$-norm quantum coherence (a), QD (b), concurrence (c), QS (d),  BNL (e), and average fidelity (f) versus temperature $T$ at $J_x=0.8$, $J_y=0.5$, $J_z=0.3$, $\theta=\pi/4$, and $\mathcal{D}_z=0$ for different fixed values of $\Gamma_z$.}
\label{f4}
\end{figure*}
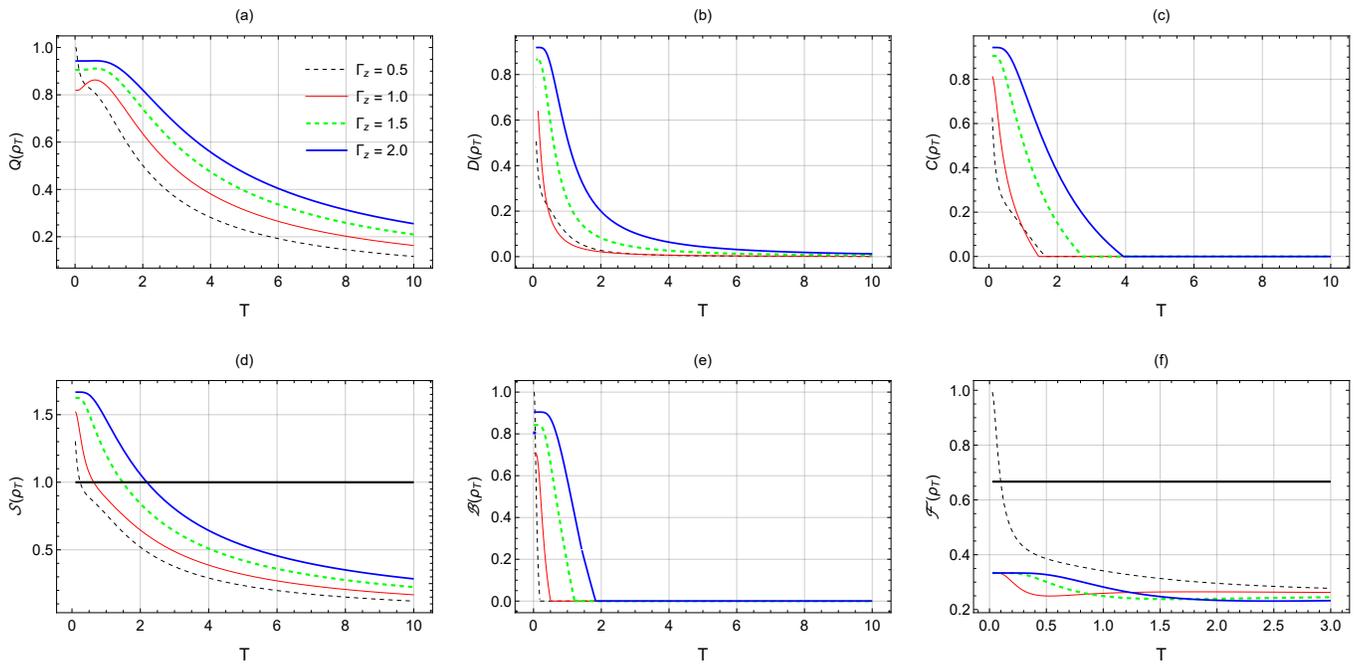

\subsection{Impact of DM Interaction}
Next, we examine the impact of different fixed values of DM interaction, without introducing KSEA interaction, on the quantum resources and the average fidelity of teleportation. This analysis is presented in Fig. \ref{f3} where the considered functions are evaluated against temperature $T$ for various fixed values of DM interaction: $\mathcal{D}_z = 0.5$ (dashed black), $\mathcal{D}_z = 1.0$ (solid red), $\mathcal{D}_z = 1.5$ (dashed green), and $\mathcal{D}_z = 2.0$ (solid blue).

We observe that although DM interaction does not increase the peak values of these functions, it aids in maintaining high peak values of quantum resources even at significantly higher temperatures (compare with Fig \ref{f2} when $\mathcal{D}_z = 0$). Thus, we can consider its effect as a positive influence on all quantum coherence and correlations of the quantum channel. For instance, when $\mathcal{D}_z = 0.5$,  BNL diminishes to zero around $T \approx 0.9$, whereas for $\mathcal{D}_z = 2.0$, it decreases to zero at $T \approx 2.5$, marking more than a five-fold improvement in the sustainability of  BNL. A similar trend is observed for QD and concurrence, with even higher $T$ values. Notably, among all quantum resources,  BNL is the most delicate, followed by QS.

Despite achieving maximal entanglement at near-zero temperatures, the average teleportation fidelity remains greater than the classical limit of $2/3$, indicating successful but not maximal teleportation. At low temperatures, the value of teleportation fidelity slightly decreases with larger DM interactions. However, this underscores that the maximal entanglement in a mixed-state situation does not always guarantee ideally successful teleportation, even at the lowest $T$ values. Thus, increasing the DM interaction decreases the peak value of the average fidelity of teleportation while making it stable at higher temperatures. Nevertheless, this shows that maintaining the nonclassical characteristics at higher temperatures with increased DM interaction does not guarantee enhanced values of average teleportation fidelity.

In summary, increasing DM interaction generally enhances all the quantum correlations and coherence in our work but does not significantly increase the peak value of average teleportation fidelity, although it helps maintain these values at higher $T$. Again, the hierarchy of quantum resources, as depicted in Fig. \ref{f1}, is held.
\subsection{Impact of KSEA Interaction}
Finally, we explore the impact of various fixed values of KSEA interaction, without introducing DM interaction, on quantum resources and the average fidelity of teleportation. This analysis is illustrated in Fig. \ref{f4} when we have plotted all functions against $T$ for different fixed values of KSEA interaction: $\Gamma_z = 0.5$ (dashed black), $\Gamma_z = 1.0$ (solid red), $\Gamma_z = 1.5$ (dashed green), and $\Gamma_z = 2.0$ (solid blue) with zero value of $\mathcal{D}_z$.
One can observe that, unlike DM interaction, KSEA interaction increases the peak values of these measures [compare with Fig \ref{f3} when $\Gamma_z = 0$]. Besides, it helps to maintain high peak values of quantum coherence and quantum correlations even at significantly higher temperatures. In general, its effect is positive on all quantum resources of the quantum channel. For instance, when $\Gamma_z = 0.5$,  BNL diminishes to zero around $T \approx 0.2$, whereas for $\Gamma_z = 2.0$, it decreases to zero at $T \approx 1.8$, marking a more than nine-fold improvement in the sustainability of  BNL. A similar trend is observed for QD and concurrence. Again, among all quantum resources,   BNL remains the most delicate, followed by QS.

Despite achieving maximal entanglement at very low temperatures, the average teleportation fidelity only remains greater than the classical limit ($2/3$) when KSEA interaction is very small [look at Fig. \ref{f4}(f)]. Namely, at $\Gamma_z = 0.5$, the average fidelity of teleportation is greater than $2/3$ at near-zero temperatures, whereas increasing KSEA interaction does not help in achieving fidelity greater than $2/3$ even at low temperatures. This shows that increasing the KSEA interaction does not enhance teleportation fidelity and can not guarantee successful teleportation. { This observation aligns with findings in other studies, such as \cite{nandi2018two}, which indicates that teleportation fidelity is not solely determined by entanglement measures such as concurrence and purity. Specifically, even states with higher concurrence and purity may exhibit lower teleportation fidelity due to the influence of other nonlocal properties that are not captured by these traditional metrics. Therefore, our results further emphasize that increasing KSEA interaction, despite enhancing entanglement, does not guarantee successful teleportation, as the fidelity is also influenced by additional state parameters beyond concurrence and purity.}

Although a large value of KSEA interaction does not enhance the average fidelity of teleportation, it is beneficial for enhancing quantum coherence and correlations (quantum resources). Therefore, increasing KSEA interaction enhances all the correlations and coherence but does not increase the peak value of average teleportation fidelity. While it helps maintain these values at higher $T$. Overall, the impact of KSEA interaction on teleportation fidelity is negative.

Achieving a quantum advantage for teleportation is unsuccessful across all temperatures with different fixed KSEA interaction values. Although quantum coherence and correlations generally persist at nonzero levels for even higher temperatures, indicating the presence of BNL, entanglement, discord, and steering, successful quantum teleportation (greater than $\mathcal{F}(\rho_T)=2/3$) is not achieved in this scenario. Additionally, we find that the channel state exhibits BNL and steerability, as well as entanglement, yet teleportation remains unachievable.

{ \section{Conclusion and outlook}\label{sec6}
In this study, we conducted a comprehensive comparative analysis of various coherence and correlation measures, including Bell nonlocality, quantum steering, quantum entanglement, quantum discord, and $l_1$ norm of quantum coherence, within the context of a two-qubit Heisenberg XYZ model under the Dzyaloshinsky-Moriya (DM) and Kaplan–Shekhtman–Entin-Wohlman–Aharony (KSEA) interactions with variable Zeeman splitting. We treated the Gibbs thermal state of this model as a quantum channel for teleportation, evaluating the average teleportation fidelity. These coherence and correlation measures, along with average teleportation fidelity, were systematically examined as functions of temperature while varying key channel parameters such as the Zeeman splitting, KSEA interaction, and DM interaction.

Our findings indicate that the presence of entanglement in the context of the Heisenberg spin channel does not inherently guarantee successful teleportation, particularly in mixed-state scenarios. This conclusion not only aligns with but also extends the insights from previous works, including the study by Nandi \textit{et al.} \cite{nandi2018two} on two-qubit X-states. There are other proposals that somehow patronize our study by Adhikari \emph{et al.} \cite{adhikari2008teleportation}.
We have shown that applying a magnetic field to both spins instead of one spin significantly enhances quantum coherence and teleportation fidelity. However, these enhancements deteriorate with increasing temperature due to thermal fluctuations, underscoring the critical role of temperature in degrading quantum properties.

Moreover, our results indicate that while increasing the DM interaction maintains higher peak values of quantum resources at elevated temperatures, it does not significantly boost the average teleportation fidelity, which remains above the classical threshold of $2/3$. This finding underscores that maximal entanglement in mixed states does not guarantee successful teleportation, corroborating Nandi \textit{et al.}'s \cite{nandi2018two}  conclusions that entanglement alone cannot ensure teleportation success. Similarly, while both DM and KSEA interactions generally increase the strength of all coherence and correlations, they fail to provide significant support for the teleportation process.
Also, our analysis reveals that increasing KSEA interaction fails to improve teleportation fidelity, echoing Nandi \textit{et al.}'s \cite{nandi2018two} assertion that higher concurrence and purity do not always translate into improved teleportation fidelity. Besides,
Nandi \textit{et al.} \cite{nandi2018two} emphasized the detrimental effects of noise and decoherence on teleportation fidelity in two-qubit X-states, highlighting that even with maximal entanglement, the fidelity is compromised by the mixed nature of the states involved.
Our study aligns with this perspective, demonstrating that despite the presence of maximal entanglement, the mixed-state nature impairs teleportation fidelity.

We also scrutinized various quantum resources within the traditional hierarchy of quantum resources and found that this hierarchy remains valid in our model. Specifically, Bell nonlocality emerges as the most fragile correlation, while the $l_1$-norm of coherence stands out as the most robust indicator, capable of capturing all forms of quantum coherence. This refined understanding of quantum resources provides a more comprehensive framework for analyzing teleportation fidelity in mixed-state scenarios.
}

\section*{Acknowledgements}
M.G. and S.H. were supported by Semnan University under Contract No. 21270.

\section*{Author Contributions}
A.A., M.T.R., and S.H. contributed to the writing of the manuscript and played a role in software development and conceptualization. S.A. and S.H. provided supervision throughout the process and carefully reviewed the manuscript. M.G. and H.A revised final draft of the manuscript.

\section*{Disclosures}
The authors declare that they have no known competing financial interests.

\section*{Data availability}
No datasets were generated or analyzed during the current study.

\bibliography{bibliography}

\end{document}